\documentstyle[eqsecnum,prd,aps,floats]{revtex}
\begin{document}

\newcommand{\nl}{\nonumber \\}
\newcommand{\eq}[1]{Eq.~(\ref{#1})}
\newcommand{\npol}{\bigtriangleup}
\newcommand{\dpol}{\bigtriangledown}
\newcommand{\cstar}{{\bf C}^\ast}

\preprint{\vbox{\hbox{CALT-68-2147}
                \hbox{hep-th/9712011}  }}
 
\title{Exact solutions for some ${\cal N}=2$ supersymmetric
$SO(N)$ gauge theories with vectors and spinors}
 
\author{Mina Aganagic and Martin Gremm}
 
\address{California Institute of Technology, Pasadena, CA 91125}

\maketitle

\begin{abstract}
We find exact solutions for ${\cal N}=2$ supersymmetric
$SO(N)$, $ N=7,9,10,11,12$ gauge theories with
matter in the fundamental and spinor representation.
These theories, with specific numbers of vectors and spinors, arise
naturally in the compactification of type IIA string theory on suitably
chosen Calabi--Yau threefolds. Exact solutions are obtained by using
mirror symmetry to find the corresponding type IIB compactification. 
We propose generalizations of these results to cases with arbitrary
numbers of massive vectors and spinors.
\end{abstract}

\section{Introduction}
Over the last few years our understanding of the low energy behavior of 
supersymmetric field theories has advanced substantially. Using field theory
arguments, Seiberg and Witten found the exact solution of ${\cal N}=2$ supersymmetric
$SU(2)$ gauge theories with matter in the fundamental representation \cite{sw}.
Subsequently many other gauge groups with fundamental matter
were analyzed in this way \cite{curves,all,so2n,so2n1}.  It turned
out to be rather difficult to generalize these results to theories with 
matter in any other representation because  in these cases the curves encoding
the gauge coupling are usually not hyperelliptic. General
Riemann surfaces have more parameters than can be fixed by studying various
limits of the gauge theory. 

String theory offers a way to construct supersymmetric gauge theories 
geometrically. In this approach, the curves encoding the gauge couplings
are real physical objects, which can be determined 
by string theory arguments. This provides a way to find exact solutions of
gauge theories for which the curves are not hyperelliptic. 

There are two methods of constructing four--dimensional
gauge theories from string theory.
The first uses configurations of branes in type IIA string theory
\cite{branes1,branes2}. Lifting the type IIA construction to M-theory, one
can obtain the exact solution to the gauge theory living on the common 
directions of the branes. This approach allows the elegant
construction of theories with matter in the fundamental or two-index tensor
representation. However, at present there are no brane constructions for
theories with higher index tensors or spinors of $SO(N)$.

The second approach involves compactifying type IIA string theory on 
$K3$ fibered Calabi--Yau threefolds
\cite{hetdual,ftlimit,periods,brodie,6a,cand}.
There is a clear physical picture of how a theory with a gauge group
corresponding to the singularity type \cite{arnold} of the $K3$ arises in
four dimensions. This provides a way to construct
a large class of $d=4$, ${\cal N}=2$ gauge theories. Type IIA string theory,
compactified on such a Calabi--Yau threefold, is conjectured
to be dual to heterotic strings on $K3\times T^2$, where the $E_8\times E_8$
gauge group is broken by instantons \cite{hetdual}.
One can use the  breaking patterns of the
$E_8\times E_8$ adjoint to determine the charged matter content of the gauge
theory \cite{matter}.
At tree level this gives a complete description of the gauge theory in four
dimensions. However, there are quantum corrections to the tree level results
due to world sheet instantons in the type IIA theory. These instanton
corrections can be summed up using mirror symmetry \cite{mirror} (for
reviews see \cite{reviews}).
Mirror symmetry pairs up two different Calabi--Yau manifolds such that type IIA
compactified on the first gives rise to the same string world sheet theory,
and therefore to the same gauge theory
as type IIB compactified on the other.
We obtain exact field theory results by studying the properties of the type IIB
Calabi--Yau, because there are no quantum corrections to the tree level results
in the type IIB compactification.

Since the description of the gauge theory depends only on the local properties
of the Calabi--Yau, it is sufficient to consider local approximations 
to the threefolds. Constructing these without first
constructing the entire Calabi--Yau is called `Geometric Engineering'
\cite{engineer,engineer2}.
We will not use this approach for the analysis in this paper, because global
descriptions of the Calabi--Yau manifolds we will consider are
available.

The approach of constructing field theories from string theory compactifications
may be more indirect than the brane picture, but it has the advantage of
providing descriptions
of theories with matter in higher index tensor representations, and in the
spinor representation for $SO(N)$ theories \cite{6a,cand,matter}.
In this paper we analyze $SO(N)$,
$N=7,9,10,11,12$  theories with vectors and spinors.  We find exact
solutions on the Coulomb branch of these theories in the form of ALE
fibrations over a sphere.

So far, no theories of this type have been analyzed in the brane picture, but
the solutions we present here may turn out to be
useful in constructing appropriate brane configurations. 
The fact that our curves agree with known field theory
results, in the cases where these are available, lends further support to
the conjectured duality between type IIA compactified on a Calabi--Yau and
heterotic strings on $K3\times T^2$. It also provides additional examples
where the instanton series can be summed up correctly, using the
mirror map.

The paper is organized as follows: In the first two subsections of Sec.~\ref{ii}
we review the construction of
a class of Calabi--Yau threefolds which give rise to $d=4$, ${\cal N}=2$
$SO(10)$ and $SO(12)$ gauge theories with specific numbers of
fundamentals and spinors.
We use the toric description of these manifolds 
to find explicit expressions for the mirror manifolds. 
A local approximation to the mirror manifolds in the form of ALE fibrations
provides the exact solutions for these theories. We also propose 
generalizations to arbitrary numbers of massive vectors and spinors and
perform several consistency checks on our results.
The non--simply laced cases $SO(7)$, $SO(9)$ and $SO(11)$ are the subject
of the next three subsections. In these cases we slightly modify the
conventional method of finding the mirror to obtain the exact solutions in the 
most convenient form. These modifications are explained in Sec.~\ref{iic}.
We summarize our results in Sec.~\ref{iii}.

\section{ Exact solutions from mirror symmetry}
\label{ii}

Type IIA string theory, compactified on a Calabi--Yau
threefold that is both an elliptic and a $K3$ fibration, gives rise to an
${\cal N}=2$ gauge theory in four dimensions. Such elliptically
fibered manifolds are defined by
\begin{equation}
y^2 = x^3 + xf(z_1,z_2)  + g(z_1,z_2),
\label{fibercy}
\end{equation}
where $f$ and $g$ are functions of the base coordinates $z_1,\,z_2$. 
For this equation to define a Calabi--Yau, the functions $f$ and $g$ must
be of the form
\begin{eqnarray}
\label{fg}
f(z_1,z_2) &=& \sum_{i=0}^I z_1^{8-i} f_{8+n(4-i)}(z_2) \nl
g(z_1,z_2) &=& \sum_{j=0}^J z_1^{12-j} f_{12+n(4-j)}(z_2) ,
\end{eqnarray}
where the subscript on the polynomials $f$ and $g$ in the sums indicates their
degree in $z_2$. $I$ and $J$ are the maximum values of $i$
and $j$ such that the degree is not negative. 
We can view this threefold as an elliptic fibration over the
Hirzebruch surface $F_n$ or as a $K3$ fibration over a sphere parameterized
by $z_2$.

Type IIA string theory compactified on this Calabi--Yau is conjectured to 
be dual to heterotic strings compactified on $K3\times T^2$ with $12-n$ and
$12+n$ instantons embedded in the first and second $E_8$ of the $E_8\times E_8$
gauge group \cite{hetdual,ftheory} and all Wilson lines switched off.
The coefficients of the monomials in \eq{fibercy} that are
proportional to $xz_1^4$ and $z_1^6$ correspond to the moduli of the $K3$ and
the other terms specify the $E_8\times E_8$ gauge bundle. The coefficients of
terms with lower powers of $z_1$ define the embedding of $12-n$ instantons
in the first $E_8$ and the remaining terms do the same for the $12+n$ 
instantons in the second $E_8$ \cite{ftheory}.

For generic choices of the polynomials $f$ and $g$, the instantons break the
$E_8\times E_8$ gauge group of heterotic strings as far as possible. The
$E_8$ with $12+n$ instantons is broken completely (for $n\ge 0$) while the
other is broken to some terminal group without matter. This is the case
that was studied in \cite{hetdual,ftlimit,periods,brodie} for various instanton embeddings.

Here we consider more restrictive instanton embeddings, which result in 
larger unbroken subgroups of the $E_8$ with $12-n$ instantons. 
On the type IIA side such instanton embeddings
correspond to choosing Calabi--Yau threefolds that have a more severe singularity
in their $K3$ fiber than one would get from the generic choice of polynomials.
For example, we can consider the Calabi--Yau defined by setting
\begin{eqnarray}
\label{cond}
f_{8-2n}&=&h_{4-n}^2 \nl
g_{12-3n}&=&h_{4-n}^3 \nl
g_{12-2n}&=&q_{6-n}^2-f_{8-n}h_{4-n}
\end{eqnarray}
and choosing the coefficients of lower powers of $z_1$ to vanish. $h_{4-n}$ and
$q_{6-n}$ are polynomials in $z_2$ of the degree indicated by the subscripts.
One can use Kodaira's classification to determine the singularity type of the
$K3$ fiber. The definitions above ensure that the fiber has a split $D_5$
singularity \cite{6a}. 
We can make this manifold smooth by blowing up a collections of 
spheres in the base of the $K3$, i.e., by modifying its K\"ahler structure.
The intersection forms of these spheres give the entries in the Cartan matrix
of the corresponding gauge group ($SO(10)$ for $D_5$). Compactifying type IIA
on a Calabi--Yau with this blown--up $K3$ as a fiber results in a $d=4$ $SO(10)$
gauge theory, where the $SO(10)$ is broken to its Cartan subalgebra. This
situation arises because the 2-branes of type IIA can wrap around the blow--up
spheres with two different orientations, giving rise to a pair of $W^\pm$
bosons with a mass proportional to the area of the spheres. Shrinking a
sphere to zero size makes the $W^\pm$ massless, which corresponds to 
unhiggsing an $SU(2)$ factor. Since the blow--up spheres have intersection
forms determined by the singularity type, the corresponding $SU(2)$ factors
link up to make the gauge group indicated by the singularity type.
Thus it is clear that the K\"ahler structure moduli are related to the
coordinates on the Coulomb branch of the $d=4$ gauge theories in the type
IIA picture. On the heterotic side, these blow--ups correspond to switching
on Wilson lines to break the gauge group.

The moduli space of a Calabi--Yau is the space of all possible choices of its
K\"ahler and complex structure.
Locally, it is a direct product of the complex and K\"ahler structure moduli
spaces. In type IIA compactifications, the K\"ahler moduli are vector
multiplets while the complex structure moduli are hypermultiplets of the
space--time theory. Since the
dilaton is also a hypermultiplet, the K\"ahler moduli space does not have 
perturbative string corrections. However, there are world sheet instanton
corrections to the K\"ahler moduli space, which are related
to gauge theory instantons via the duality to heterotic strings \cite{engineer}.
Mirror symmetry provides a way to sum up these corrections.

To find the mirror manifold of the type IIA Calabi--Yau it is convenient to
encode its salient properties using toric geometry \cite{mirror,reviews,math}.
For the cases we are considering here this was worked out
in \cite{6a,cand}, so we will only summarize the results.
The natural starting point for the application of toric methods is the 
representation of  the Calabi--Yau as a hypersurface in a weighted
projective space.  The toric data
consist of two polyhedra, the Newton polyhedron $\npol$, and the dual
polyhedron $\dpol$. The vertices of $\dpol$ are the normal vectors on the
facets of $\npol$ and vice versa. 
The Newton polyhedron encodes the complex structure
of the manifold. For every monomial that appears in the defining equation
of the manifold, the vector of exponents gives a corresponding point in
the polyhedron. Some of the vertices of the dual polyhedron define the
ambient space in which the
Calabi--Yau is embedded and the others encode the K\"ahler structure of the
blow--up spheres in the base of the $K3$. 

The role of the two polyhedra $\npol$ and $\dpol$ are exchanged under mirror
symmetry. In order to construct the manifold for type IIB compactifications,
we take the dual polyhedron, $\dpol$, to encode the complex structure of the
mirror Calabi--Yau and the Newton polyhedron, $\npol$, to define the toric
variety in which it is embedded. The vertices of the dual polyhedron that
encode the K\"ahler structure of the blow--ups on the type IIA side determine
the complex structure of the type IIB manifold. 

On the type IIB side, the complex structure moduli are vector multiplets
and the K\"ahler structure and the dilaton are hypermultiplets. As on the
type IIA side, the vector moduli space is not corrected by perturbative 
string effects but on the type IIB side the world sheet instanton corrections
are absent as well \cite{engineer}. Thus the classical description of the
complex structure moduli space of the type IIB Calabi--Yau is exact.
Since these moduli encode the behavior of the gauge theory, we can read off
the exact solutions from the IIB manifold.

Below, we discuss a series of Calabi--Yau manifolds
that give rise to $SO(10),\,SO(12)$ and $SO(7),\,SO(9)$ and $SO(11)$
gauge groups with spinors and fundamentals.
In the first two cases we find exact solutions using
Batyrev's construction of the mirror \cite{math}.
For the non--simply laced cases we slightly modify the construction to simplify
the resulting curves. These modifications are explained in Sec.~\ref{iic}.

\subsection{The Calabi--Yau for $SO(10)$}
\label{iia}

The dual polyhedron for the Calabi--Yau that gives rise to an $SO(10)$ gauge
theory with $4-n$ spinors and $6-n$ vectors was constructed in
\cite{6a}\footnote{Note that in Refs.~\cite{6a,cand} the unhiggsing of the
$E_8$ with $12+n$ instantons was studied while we are unhiggsing the $E_8$
with $12-n$ instantons.}.
The derivation there uses Tate's algorithm and a more general
form of the defining equation, \eq{fibercy}, that makes it easier to encode
the split or nonsplit property of the singularity. The same polyhedron was 
also found in \cite{cand}, using toric arguments only. Using the basis
of \cite{cand}, the dual polyhedron, $\dpol$, is given by the vertices
\begin{equation}
\begin{tabular}{lll}
$\tilde{v}_1 = (-1, 0, 2, 3) $ &
$\tilde{v}_2 = (1, -n, 2, 3) $ &
$\tilde{v}_3 = (0, -1, 2, 3) $ \\
$\tilde{v}_4 = (0, 0, -1, 0) $ &
$\tilde{v}_5 = (0, 0, 0, -1) $ &
$\tilde{v}_6 = (0, 0, 0, 0) $ \\
$\tilde{v}_7 = (0, 0, 2, 3) $ &
$\tilde{v}_8 = (0, 1, 2, 3) $ &
$\tilde{v}_9 = (0, -2, 2, 3) $ \\
$\tilde{v}_{10} = (0, -2, 1, 2) $ &
$\tilde{v}_{11} = (0, -1, 1, 1) $ &
$\tilde{v}_{12} = (0, -1, 0, 1) $ \\
$\tilde{v}_{13} = (0, -1, 0, 0) $.
\end{tabular}
\end{equation}
This list of vertices includes all points that do not lie on codimension one
facets of the dual polyhedron, i.e., this polyhedron encodes a fully blown--up
type IIA manifold. 
The vertices $\tilde{v}_1,\ldots,\tilde{v}_8$ define the toric variety in
which the type IIA manifold is embedded and the remaining vertices correspond
to the blow--up spheres needed to repair the $D_5$ singularity.
The vertices of the corresponding Newton polyhedron are
\begin{equation}
\begin{tabular}{lll}
$v_1=(2,1,-1,1) $ & $ v_2=(3,1,1,0)$ & $v_3=(0,0,1,-1)$ \\
$ v_4=(6,1,1,1) $ & $ v_5=(0,0,-2,1) $ & $v_6=(6,-6,1,1)$\\
$ v_7=(-6-6n,-6,1,1) $ & $ v_8=(n-6,1,1,1) $ & $v_9=(n-3,1,1,0)$\\
$ v_{10}=(n-2,1,-1,1). $ &&\\
\end{tabular}
\end{equation}
Note that for $n=4$ the vertices $v_1$ and $v_{10}$ become identical which
allows us to drop one of them.

We can use the information encoded in the dual pair of polyhedra, $\npol , \,
\dpol$, to construct the mirror manifold of our initial Calabi--Yau. Batyrev's
construction of the mirror \cite{math} requires that we switch the roles of
the two polyhedra. An embedding polynomial defining the mirror manifold is
given by
\begin{equation}
W = \sum_j a_j \prod_i x_i^{v_i \cdot \tilde{v}_j + 1} = 0,
\label{hyper}
\end{equation}
where the $x_i$ are coordinates in a weighted projective space
(or more generally in a toric variety). In the cases we consider here there
are nine or ten vertices in the Newton polyhedron, corresponding to the 
same number of coordinates in the hypersurface constraint, \eq{hyper}.
We can eliminate some of these coordinates using the $\cstar$ actions that 
define the identifications of coordinates in the embedding space. Sets of
weights for the $\cstar$
actions can be found by looking for sets of five vertices in $\npol$ such
that
\begin{equation}
\sum_i v_i k_i = 0,
\end{equation}
where the coefficients satisfy $k_i \ne 0$. One can use these $\cstar$ actions
to set all but five of the coordinates in \eq{hyper} to one. This gives
a description of the Calabi--Yau in some local coordinate patch with one
remaining $\cstar$ action. 
For our purposes it is most convenient to retain $x_1,x_2,x_3,x_6,x_7$
 and set the remaining coordinates to one. This amounts to choosing a patch
in which the relevant properties of the Calabi--Yau are described most easily.
Using these coordinates we find  the following
defining equation for the mirror manifold
\begin{eqnarray}
W &=&  x_7^{12+6n}+a_0 x_1^{4-n}x_2^{6-n}x_6^{12+6n} +
	a_1 x_1 x_2^2(x_6 x_7)^{12}
	+ a_2 x_1^2 + a_3 x_2 x_3^2 + a_4 x_1x_2x_3(x_6x_7) \nl
	&& + a_5 x_1^2 x_2^3 (x_6x_7)^6+a_6 x_1^3x_2^4+a_7x_2(x_6x_7)^{18}+
	a_8(x_6x_7)^{16}+a_9x_2x_3(x_6x_7)^9\\
	&&+a_{10}x_1(x_6x_7)^8+ a_{11}x_3(x_6x_7)^7.\nonumber
\end{eqnarray}
This Calabi--Yau is
a $K3$ fibration. We can make this explicit by defining $x_0=x_6x_7$ and
$\zeta= (x_7/x_6)^{6+3n}$. Using the freedom to rescale $x_1,x_2$ and $x_3$
to eliminate three of the coefficients $a_i$ we obtain
\begin{eqnarray}
\label{k3fiber}
W &=&  \left(\zeta+a_0 \frac{x_1^{4-n}x_2^{6-n}}{\zeta}\right)x_0^{6+3n} 
	 -2x_1 x_2^2x_0^{12} - x_1^2 + x_2 x_3^2 + a_4 x_1x_2x_3x_0 \\
	&& + a_5 x_1^2 x_2^3 x_0^6+a_6 x_1^3x_2^4+a_7x_2x_0^{18}+
	a_8x_0^{16}+a_9x_2x_3x_0^9 +a_{10}x_1x_0^8+ a_{11}x_3x_0^7.\nonumber
\end{eqnarray}
The first term in this equation describes the base sphere and the remaining
terms define a $K3$.
Approximating the $K3$ locally as an ALE space, we can bring this expression
into a form that is equivalent to a Seiberg-Witten curve.
In order to do this, we set $x_0=1$ and
observe that the first three terms in the $K3$ part give a
three--coordinate form of a $D_5$ singularity located at the origin.
The terms with coefficients $a_5$ and $a_6$ are irrelevant near the
singularity and can be neglected for our present purposes. The remaining
terms are the versal deformations of the $D_5$ singularity. 
The following chain of substitutions brings the singularity into the standard
form:
\begin{eqnarray}
x_3 &=& y - \frac{1}{2}\left( a_9+a_4x_1\right) \nl
a_8 &=& c_1 + \frac{1}{16}
	\left( 8 a_{11} a_9+4a_{11}a_{10}a_4-4a_{10}^2-a_{11}^2a_4^2\right)\nl
a_7 &=& c_2+\frac{a_9}{8}\left(2a_9+2a_{10} a_4-a_{11} a_4^2\right) \\
a_{10} &=& -c_3 +\frac{a_4}{16}\left(8a_{11}+a_4 a_9^2\right)\nl
a_9 &=& \frac{2}{a_4} c_4\nl
a_{11} &=& -2 (-c_0)^{1/2}. \nonumber
\end{eqnarray}
Neglecting an irrelevant term proportional to $x_1^2 x_2$ we obtain the
standard form of the $D_5$ singularity after shifting 
\begin{equation}
\label{x1}
x_1 = x - \frac{1}{8}\left(  4c_3-c_4^2+4c_4z+8z^2\right)
\end{equation}
and defining $z=x_2$: 
\begin{equation}
\label{sing}
W =  \left(\zeta+a_0 \frac{x_1^{4-n}z^{6-n}}{\zeta}\right) -x^2+z^4+y^2z 
	-2(-c_0)^{1/2}y+c_4z^3+c_3z^2+c_2z+c_1+\cdots,
\end{equation} 
where  \eq{x1} should be substituted for $x_1$.
The ellipsis denotes contributions from terms that are irrelevant close to the
singularity.
Neglecting these terms amounts to switching off gravity or conversely taking
the field theory limit \cite{ftlimit,periods}.
The strange choice for the redefinition of $a_{11}$ will become clear below.

This expression is equivalent to a Seiberg-Witten curve for $SO(10)$ with
$6-n$ fundamentals and $4-n$ spinors. The coefficients $c_0,\ldots,c_4$ are
the gauge invariant coordinates on the moduli space and $a_0$ can be 
interpreted as the strong coupling scale of the gauge theory,
$a_0=\Lambda^{2\beta_0}$. The beta function for this $SO(10)$ theory is given by
$\beta_0 = 8-N_f-2N_s$.

The general method for converting $D_n$ type ALE fibrations into 
Seiberg-Witten curves was first introduced in \cite{so2n} to find the
curves for $SO(2N)$ gauge groups without matter. Using the same approach
we integrate out $y$ from \eq{sing} and multiply by $z$. Absorbing a factor
of $z$ into $\zeta$ gives
\begin{equation}
\label{so10nice}
W =  \left(\zeta+\Lambda^{2\beta_0}
	 \frac{x_1^{4-n}z^{6-n+2}}{\zeta}\right) -x^2+2P(z),
\end{equation}
where $P(z)$ is given by
\begin{equation}
P(z) = \frac{1}{2}\left(z^5+c_4z^4+c_3z^3+c_2z^2+c_1z + c_0 \right).
\end{equation}	
For $n=4$, $x$ appears only quadratically and can be integrated out
trivially.  The substitutions $\zeta=y-P(z)$ and $z\to z^2$ result in a
double cover version of the curve for $SO(10)$ with two fundamentals
\begin{equation}
y^2=P^2(z^2)-\Lambda^{12}z^8.
\end{equation}
Note that for the asymptotically free cases, $n=2,3,4$,
both $x$ and $y$ appear at most quadratically and can be integrated out.
In the cases with one or two spinors of $SO(10)$, $n=2,3$, we still obtain a
curve but it is no longer hyperelliptic.
The $U(1)$ gauge couplings on the Coulomb branch are encoded in the
normalized period matrix of this curve. 
The Seiberg-Witten 1-form needed to evaluate the period matrix, can be derived
from the unique holomorphic 3-form, $\Omega$, of the original Calabi--Yau
\cite{periods}.

It is very tempting to modify \eq{sing} to allow an arbitrary number of massive
spinors and vectors. This can probably be achieved by replacing the fibration
over the sphere in \eq{sing} according to
\begin{equation}
\label{so10curve}
\zeta+a_0\frac{1}{\zeta} x_1^{4-n}z^{6-n} \to
	\zeta+a_0\frac{1}{\zeta}\prod_{i=1}^{N_s}(x_1-m^4_i)
		\prod_{j=1}^{N_f}(z-m_j^2),
\end{equation}
where the $m_i$ are the masses of the $N_s$ spinors and the $m_j$ are the masses
of the $N_f$ vectors. Using \eq{so10curve} and substituting $\zeta=y-P(z)$,
$z\to z^2$ in \eq{so10nice}, we get
\begin{equation}
\label{guess}
y^2= x^2\left( y-P(z^2)\right)+ P^2(z^2) - 
\Lambda^{2\beta_0}z^4\prod_{i=1}^{N_s}(x_1-m^4_i)\prod_{j=1}^{N_f}(z^2-m_j^2).
\end{equation}
The normalized period matrix of this surface encodes the gauge couplings
on the Coulomb branch. Here, there is no natural 2-form inherited from
$\Omega$, because \eq{guess} is generally
not a parametrization of a local approximation
to a Calabi--Yau. To compute the gauge couplings from this surface, one needs
to identify the 2-cycles and construct a suitable 2-form directly.

Our proposal, \eq{so10curve}, ensures plausible behavior when either a spinor
or a vector is integrated out. Integrating out a vector and a spinor at
the same time, we can flow between the theories we obtained from mirror
symmetry. To check our solution further, we consider breaking the $SO(10)$
gauge group to $SO(8)\times U(1)$ by giving a large VEV, $M$, to one component
of the $SO(10)$ adjoint. Under this breaking the fundamentals decompose into
fundamentals of $SO(8)$ and singlets with $U(1)$ charge. The spinors decompose
as ${\bf 16}\to {\bf 8}_c^1\oplus {\bf 8}_s^{-1}$, where the superscripts
denote the $U(1)$ charge \cite{slansky}. Both the singlets and the two 
spinor representations of $SO(8)$ acquire a large mass and should drop
out from our solution. Taking $M$ to infinity, the piece proportional
to $c^2_4\approx M^4$ will dominate \eq{x1}. Replacing $x_1$ by $M^4$,
rescaling \eq{so10nice} by appropriate powers of $M$ and integrating out
$x$ reduces it to the $SO(8)$ curve with vector matter only.

\subsection{The Calabi--Yau for $SO(12)$}

The analysis of the previous subsection can be repeated for $SO(12)$
with $r$ half hypermultiplets in the $\bf{32}$, $(4-n-r)$ half hypermultiplets
in the $\bf{32^\prime}$ and $8-n$ fundamentals. The
restrictions on the polynomials $f$ and $g$ in \eq{fg} are more complicated
for $SO(12)$ than for $SO(10)$ \cite{6a}, partly because one has
the freedom to trade matter
fields in the $\bf{32}$ for fields in the $\bf{32}^\prime$ representation. 
However, the curve of the $SO(12)$ theories depends only on the total
number of fields
in the $\bf{32}$ and $\bf{32}^\prime$, so we will drop this distinction here.
Using the vertices of the dual polyhedron given in \cite{cand},
\begin{equation}
\begin{tabular}{lll}
$\tilde{v}_1 = (-1, 0, 2, 3) $ &
$\tilde{v}_2 = (1, -n, 2, 3) $ &
$\tilde{v}_3 = (0, -1, 2, 3) $ \\
$\tilde{v}_4 = (0, 0, -1, 0) $ &
$\tilde{v}_5 = (0, 0, 0, -1) $ &
$\tilde{v}_6 = (0, 0, 0, 0) $ \\
$\tilde{v}_7 = (0, 0, 2, 3) $ &
$\tilde{v}_8 = (0, 1, 2, 3) $ &
$\tilde{v}_9 = (0, -2, 2, 3) $ \\
$\tilde{v}_{10} = (0, -2, 1, 2) $ &
$\tilde{v}_{11} = (0, -2, 0, 1) $ &
$\tilde{v}_{12} = (0, -1, 1, 1) $ \\
$\tilde{v}_{13} = (0, -1, 0, 0) $ &
$\tilde{v}_{14} = (0, -1, -1, 0) $,
\end{tabular}
\end{equation}
we find for the Newton polyhedron
\begin{equation}
\begin{tabular}{lll}
$ v_1=(2,1,-1,1) $ & $ v_2=(4,1,0,1)$ & $v_3=(0,0,1,-1)$ \\
$ v_4=(0,0,-2,1) $ & $ v_5=(-6,0,1,1) $ & $v_6=(6,0,1,1)$\\
$ v_7=(-6,6,1,1) $ & $ v_8=(-6-6n,-6,1,1) $ & $v_9=(n-2,1,-1,1)$\\
$ v_{10}=(n-4,1,0,1). $ &&\\
\end{tabular}
\end{equation}
In terms of $x_1,x_2,x_3,x_7$ and $x_8$ the hypersurface defining the
Calabi--Yau, \eq{hyper}, is given by
\begin{eqnarray}
\label{d6cy}
W &=&  \left(\zeta+a_0 \frac{x_1^{4-n}x_2^{8-n}}{\zeta}\right)x_0^{6+3n} 
-2x_1 x_2^3x_0^{12}-x_1^2x_2+x_3^2+a_4 x_1x_2x_3x_0\nl
&&+a_5 x_1^2 x_2^4 x_0^6 +a_6 x_1^3x_2^5+a_7x_2^2x_0^{18}+
	a_8x_2x_0^{16}+a_9x_0^{14}+a_{10}x_2x_3x_0^9\\
&& +a_{11}x_3x_0^7+a_{12}x_1x_0^6,\nonumber
\end{eqnarray}
where we defined $x_0=x_7x_8$ and $\zeta=(x_8/x_7)^{6+3n}$ and rescaled the
coordinates to eliminate the coefficients of the first three terms defining
the fiber. The terms with
coefficients $a_5$ and $a_6$ are again irrelevant near the singularity. Making
the substitutions
\begin{eqnarray}
x_3 &=& x -\frac{1}{2}\left( a_{11}+a_{10}x_2+a_4x_1x_2\right)\nl
x_1 &=& y-\frac{1}{4}\left( a_{11}a_4+a_{10}a_4x_2+4x_2^2\right)\\
x_2 &=& z \nonumber
\end{eqnarray}
and neglecting an irrelevant piece proportional to $x_1^2x_2^2$ brings \eq{d6cy}
into the form
\begin{equation}
\label{sing12}
W =  \left(\zeta+a_0 \frac{x_1^{4-n}z^{8-n}}{\zeta}\right) +x^2+z^5-y^2z +
        2(c_0)^{1/2}y+c_5z^4+c_4z^3+c_3z^2+c_2z+c_1+\cdots.
\end{equation}
In this expression, $x_1$ is given by
\begin{equation}
x_1 = y -\frac{1}{8}\left( 4c_4-c_5^2+4c_5z+8z^2\right).
\end{equation}
We can identify $a_0$ with the strong coupling scale of the $SO(12)$ gauge
theory: $a_0=\Lambda^{2\beta_0}$. The $\beta$-function for this theory is
given by $\beta_0=10-N_f-2N_s$, where $N_s$ counts the number of half
hypermultiplets in the spinor representation of $SO(12)$. One can check 
that for $n=4$ \eq{sing12} reduces to the known curve for $SO(12)$ with
four fundamentals \cite{all,so2n}.
In the asymptotically free cases, $n=2,3,4$, this expression
reduces to a curve, because both $x$ and $y$ appear at most quadratically.

Again we conjecture that \eq{sing12} can be modified to accommodate $N_s$
spinors with masses $m_i$ and $N_f$ vectors with masses $m_j$ by the following
substitution
\begin{equation}
\zeta+a_0\frac{1}{\zeta} x_1^{4-n}z^{8-n} \to
        \zeta+a_0\frac{1}{\zeta}\prod_{i=1}^{N_s}(x_1-m^4_i)
                \prod_{j=1}^{N_f}(z-m^2_j).
\end{equation}
As in the $SO(10)$ case, this results in an expression that shows the expected
behavior under adjoint breaking of the $SO(12)$ to $SO(10)$. The substitution
above also ensures that spinors and vectors can be integrated out consistently.

\subsection{The Calabi--Yau for $SO(7)$}
\label{iic}

The $SO(7)$ theory with $3-n$ fundamentals and $8-2n$ spinors differs from
the theories we considered above in several respects. It is our first
example of a non--simply laced group. Unlike in the previous cases, the
$K3$ part of the Calabi--Yau cannot have a singularity of a type that
corresponds to the gauge group, since a $K3$ can only have ADE type
singularities. Thus we should expect some mixture of fiber and base
coordinates even if there is no matter in the theory. The second difference
is that the $SO(7)$ theory makes sense
only for $n=2,3$. For $n=4$, the fiber of the type IIA manifold cannot have
a semisplit $D_4$ singularity \cite{6a}, which would give rise to an $SO(7)$
gauge theory. Thus we cannot consider the case without spinors
to compare to known results.
Apart from that, it will turn out that the most convenient representation
of the $SO(7)$ curve requires a slight modification of Batyrev's construction
of the mirror.

The polar polyhedron giving rise to the $SO(7)$ gauge theory is defined by the
vertices
\begin{equation}
\begin{tabular}{lll}
$\tilde{v}_1 = (-1, 0, 2, 3) $ &
$\tilde{v}_2 = (1, -n, 2, 3) $ &
$\tilde{v}_3 = (0, -1, 2, 3) $ \\
$\tilde{v}_4 = (0, 0, -1, 0) $ &
$\tilde{v}_5 = (0, 0, 0, -1) $ &
$\tilde{v}_6 = (0, 0, 0, 0) $ \\
$\tilde{v}_7 = (0, 0, 2, 3) $ &
$\tilde{v}_8 = (0, 1, 2, 3) $ &
$\tilde{v}_9 = (0, -2, 2, 3) $ \\
$\tilde{v}_{10} = (0, -1, 1, 1) $ &
$\tilde{v}_{11} = (0, -1, 0, 1) $ &
\end{tabular}
\end{equation}
and the corresponding Newton polyhedron is given by
\begin{equation}
\begin{tabular}{lll}
$ v_1=(4,2,0,1) $ & $ v_2=(0,0,-2,1)$ & $v_3=(0,0,1,-1)$ \\
$ v_4=(6,2,1,1) $ & $ v_5=(6,-6,1,1) $ & $v_6=(-6-6n,-6,1,1)$\\
$ v_7=(2n-6,2,1,1) $ & $ v_8=(2n-4,2,0,1) $. &\\
\end{tabular}
\end{equation}
Using \eq{hyper} and setting $x_4=x_7=x_8=1$, 
we find the defining equation of the Calabi--Yau
\begin{eqnarray}
\label{so7cy}
W &=&  \left(\zeta+a_0 \frac{x_1^{8-2n}}{\zeta}\right)x_0^{6+3n} 
+ x_1^2x_0^{12}+x_1x_2^3+x_2^2+a_4 x_1x_2x_3x_0+a_5x_1^4x_0^6\nl
&&+a_6x_1^6+a_7x_0^{18}+a_8x_3x_0^9+a_9x_2^2x_0^8.
\end{eqnarray}
The $K3$ part of this expression can be transformed into the standard form of
the classical piece of the $SO(7)$ curve using coordinate redefinitions as
in the previous subsections. This results in an expression of the form
\begin{equation}
W =  \left(\zeta+a_0 \frac{x_1^{8-2n}}{\zeta}\right)+ x^2+y^2+z^6+c_3z^4+
	c_2z^2+c_1+\cdots,
\end{equation}
where $x_1$ is some function of $x,z$ and the Casimirs $c_i$. In this format
there is no obvious way to identify the powers of the fiber coordinates that
multiply the coordinate of the lower sphere with the number of matter
fields.

This problem can be circumvented by replacing the Calabi--Yau, \eq{so7cy},
with another Calabi--Yau that encodes the same field theory information.
Recall that on the IIB side, the field theory information is encoded in the
complex structure moduli, which in turn determine the period integrals over
the three cycles of the Calabi--Yau. The K\"ahler structure moduli determine
the integrals over two cycles but do not affect the integrals over the
three cycles. Thus we can modify the K\"ahler structure of our manifold 
without changing the information about the gauge theory.

One way of seeing that the information encoded in the complex structure is
invariant under changes of the K\"ahler structure is provided by the 
$\dpol$-hypergeometric system of partial differential equations 
(see, e.g., \cite{yau} for details). The period integral over the three cycles
of the Calabi--Yau is given by
\begin{equation}
\Pi_k(a)=\int_{\gamma_k} \frac{1}{W(a,x)}\prod_p \frac{dx_p}{x_p},
\end{equation}
where $W(a,x)$ is a hypersurface constraint such as \eq{so7cy}, $x_p$ are
the coordinates of the embedding space and $a$ denotes the set of complex
structure moduli. The period integrals satisfy a set of differential 
equations
\begin{equation}
\label{rel}
{\cal D}_l \Pi_k = 0 , \quad {\cal Z}_\alpha \Pi_k = 0,
\end{equation}
where the differential operators are given by
\begin{equation}
{\cal D}_l = \prod_{l_i>0}\left(\frac{\partial}{\partial a_i}\right)^{l_i}-
	\prod_{l_i<0}\left(\frac{\partial}{\partial a_i}\right)^{-l_i}
, \quad
{\cal Z}_\alpha=\sum_i \tilde{v}_{i,\alpha} a_i \frac{\partial}{\partial a_i}
, \quad
{\cal Z}_0=\sum_i a_i \frac{\partial}{\partial a_i}+1.
\end{equation}
Here, $\tilde{v}_{i,\alpha}$ denotes the $\alpha$ component of the $i$-th
vector in the dual polyhedron and the vectors $l$ define relations between
the vertices $\tilde{v}_i$
\begin{equation}
\sum_i \tilde{v}_i l_i = 0 , \quad \sum_i l_i = 0.
\end{equation}
One can check that the hypersurface constraints obtained by Batyrev's
construction satisfy these relations. 

However, this does not exhaust the list of hypersurface constraints that 
satisfy \eq{rel}. One can find many additional manifolds by solving
these equations directly. In this approach, one does not need the information
encoded in the Newton polyhedron. This reflects the fact that all of the
information on the behavior of the gauge theory is contained in the dual
polyhedron. Different solutions to Eqs.~(\ref{rel}) will describe different 
Calabi--Yau manifolds but they will all have the same period integrals over
the three cycles and therefore they encode the same gauge theory.

We can easily find other hypersurface constraints which satisfy Eqs.~(\ref{rel})
by adding points to the Newton polyhedron that lie in its convex hull. Using
the coordinates corresponding to these points to parametrize the hypersurface
constraint guarantees that the resulting Calabi--Yau has the same period
integrals as \eq{so7cy}. Adding the vector $v_9 = (n-2,1,-1,1)$
to the Newton polyhedron and using the coordinates associated to
$v_8, v_9,v_3,v_5$ and $v_6$ we find the hypersurface constraint
\begin{eqnarray}
W &=&  \left(\zeta+a_0 \frac{x_8^{8-2n}x_9^{4-n}}{\zeta}\right)x_0^{6+3n} -
	x_8^2x_9x_0^{12}+2x_8x_9^2+x_3^2+ a_4 x_3x_8x_9x_0 \nl
	&&+a_5x_8^4x_9^2x_0^6+ a_6x_8^6x_9^3+a_7x_0^{18}+a_8x_3x_0^9
	+a_9x_9x_0^8.
\end{eqnarray}
Setting $x_0=1$, neglecting the terms with coefficients $a_5$ and $a_6$,
 and substituting
\begin{eqnarray}
x_3 &=& x - \frac{1}{2}\left( a_8+a_4x_8x_9\right) \nl
x_8 &=& y+x_9-\frac{1}{4}a_4a_8 \\
x_9 &=& z \nonumber
\end{eqnarray}
we find after redefining the complex structure parameters
\begin{equation}
\label{so7sing}
W =  \left(\zeta+a_0 \frac{x_8^{8-2n}z^{4-n}}{\zeta}\right)
	+ x^2 +z^3 - y^2z + c_2z^2+c_1z+c_0 + \cdots,
\end{equation}
where $x_8 = y+z+c_2/2$. We can identify $a_0$ with $\Lambda^{2\beta_0}$ and
the $c_i$ with the Casimirs of $SO(7)$.
Since we cannot choose $n$ to eliminate all spinors,
we cannot compare this curve directly to known results. However, higgsing
$SO(7)$ to $SO(5)$ as in Sec.~\ref{iia}, we obtain the expected curve for
$SO(5)$ with $3-n$ fundamentals.
If we modify \eq{so7sing} to allow arbitrary numbers of spinors and vectors with
arbitrary masses by replacing
\begin{equation}
\zeta+a_0\frac{1}{\zeta} x_8^{8-2n}z^{4-n} \to
        \zeta+a_0\frac{1}{\zeta}z \prod_{i=1}^{N_s}(x_8-m^2_i)
                 \prod_{j=1}^{N_f}(z-m^2_j),
\end{equation}
we can integrate out all spinors in \eq{so7sing}. Then $x$ and $y$ can
be integrated out trivially and substituting $z\to z^2$, we find the 
double cover version of the $SO(7)$ curve with $3-n$ fundamentals \cite{all,so2n1}.
Unlike in the previous cases, we can write \eq{so7sing} as a curve only for
$n=3$.

\subsection{The Calabi--Yau for $SO(9)$}

In this section we repeat the analysis of the previous sections for a
class for Calabi--Yau manifolds that lead to an $SO(9)$ gauge theory with
$5-n$ vectors and $4-n$ spinors.  The toric description of these manifolds
is given by the vertices
\begin{equation}
\begin{tabular}{lll}
$\tilde{v}_1 = (-1, 0, 2, 3) $ &
$\tilde{v}_2 = (1, -n, 2, 3) $ &
$\tilde{v}_3 = (0, -1, 2, 3) $ \\
$\tilde{v}_4 = (0, 0, -1, 0) $ &
$\tilde{v}_5 = (0, 0, 0, -1) $ &
$\tilde{v}_6 = (0, 0, 0, 0) $ \\
$\tilde{v}_7 = (0, 0, 2, 3) $ &
$\tilde{v}_8 = (0, 1, 2, 3) $ &
$\tilde{v}_9 = (0, -2, 2, 3) $ \\
$\tilde{v}_{10} = (0, -2, 1, 2) $ &
$\tilde{v}_{11} = (0, -1, 1, 1) $ &
$\tilde{v}_{12} = (0, -1, 0, 1) $ \\
\end{tabular}
\end{equation}
of the dual polyhedron. The Newton polyhedron consists of the vertices
\begin{equation}
\begin{tabular}{lll}
$v_1=(2,1,-1,1) $ & $ v_2=(6,2,1,1)$ & $v_3=(0,0,1,-1)$ \\
$ v_4=(6,-6,1,1) $ & $ v_5=(0,0,-2,1) $ & $v_6=(-6-6n,-6,1,1)$\\
$ v_7=(2n-6,2,1,1) $ & $ v_8=(n-2,1,-1,1) $. &\\
\end{tabular}
\end{equation}
Using these vectors and \eq{hyper}, we can write down the mirror. It is
convenient to
use the $\cstar$ actions to set all coordinates except $x_1,x_2,x_3,x_4$ and
$x_6$ to one.  Defining $x_0=x_4x_6$
and $\zeta=(x_4/x_6)^{6+3n}$ we get
\begin{eqnarray}
W &=& \left( \zeta+a_0\frac{x_1^{4-n}x_2^{12-2n}}{\zeta}\right) x_0^{6+3n}+
	2 x_1x_2^4x_0^{12}-x_1^2+x_3^2+a_4x_1x_2x_3x_0 \nl
&& +a_5x_1^2x_2^6x_0^6 +a_6x_1^3x_2^8+a_7x_2^2x_0^{18}+a_8x_0^{16}+
	a_9x_2x_3x_0^9+a_{10}x_1x_0^8.
\end{eqnarray}
For $SO(9)$, Batyrev's construction gives a description of the mirror in which
the matter content of the theory is visible in the fibration over the lower
sphere.  The terms with coefficients $a_5$ and $a_6$ are irrelevant near the
singularity. We can transform the fiber into the standard form for an $SO(9)$
theory by making the following substitutions
\begin{eqnarray}
x_3 &=& x-\frac{1}{2}\left(a_4 x_1x_2+a_9x_2\right) \nl
x_1 &=& y+\frac{1}{4}\left( 2 a_{10}-a_4a_9x_2^2+4x_2^4\right).\\
x_2 &=& z. \nonumber
\end{eqnarray}
Neglecting an irrelevant term of the form $x_1^2x_2^2$ and renaming the
coefficients, we find
\begin{equation}
W = \left( \zeta+a_0\frac{x_1^{4-n}z^{12-2n}}{\zeta}\right) + x^2 - y^2 +
	c_0+c_1z^2+c_2z^4+c_3z^6+z^8+\cdots,
\end{equation}
where
\begin{equation}
x_1 = y + \frac{1}{8}\left(4c_2-c_3^2+4c_3z^2+8z^4\right).
\end{equation}
It is straightforward to check that for $n=4$ this curve agrees with the curves
in \cite{all,so2n1}, once one identifies the $c_i$ with the gauge invariant
polynomials that parametrize the Coulomb branch and sets
$a_0=\Lambda^{2\beta_0}$. 

Again, the substitution
\begin{equation}
\left( \zeta+a_0\frac{x_1^{4-n}z^{12-2n}}{\zeta}\right) \to
        \zeta+a_0\frac{1}{\zeta} z^2 \prod_{i=1}^{N_s}(x_1-m^4_i)
                \prod_{j=1}^{N_f}(z^2-m_j^2)
\end{equation}
presumably results in a solution of the theory with arbitrary numbers of
massive vectors and spinors. Repeating the checks as in Sec.~\ref{iia}, we find
consistent behavior.

\subsection{The Calabi--Yau for $SO(11)$}

For $SO(11)$ with $4-n$ half hypermultiplets in the spinor representation and
$7-n$ vectors we can repeat the steps that provided the curve for $SO(7)$.
The polar polyhedron is given by the vertices
\begin{equation}
\begin{tabular}{lll}
$\tilde{v}_1 = (-1, 0, 2, 3) $ &
$\tilde{v}_2 = (1, -n, 2, 3) $ &
$\tilde{v}_3 = (0, -1, 2, 3) $ \\
$\tilde{v}_4 = (0, 0, -1, 0) $ &
$\tilde{v}_5 = (0, 0, 0, -1) $ &
$\tilde{v}_6 = (0, 0, 0, 0) $ \\
$\tilde{v}_7 = (0, 0, 2, 3) $ &
$\tilde{v}_8 = (0, 1, 2, 3) $ &
$\tilde{v}_9 = (0, -2, 2, 3) $ \\
$\tilde{v}_{10} = (0, -2, 1, 2) $ &
$\tilde{v}_{11} = (0, -2, 0, 1) $ &
$\tilde{v}_{12} = (0, -1, 1, 1) $ \\
$\tilde{v}_{13} = (0, -1, 0, 0) $ \\
\end{tabular}
\end{equation}
and the corresponding Newton polyhedron is defined by
\begin{equation}
\begin{tabular}{lll}
$v_1=(2,1,-1,1) $ & $ v_2=(6,1,1,1)$ & $v_3=(0,0,1,-1)$ \\
$ v_4=(0,0,-2,1) $ & $ v_5=(6,-6,1,1) $ & $v_6=(-6-6n,-6,1,1)$\\
$ v_7=(n-6,1,1,1) $ & $ v_8=(n-2,1,-1,1) $. &\\
\end{tabular}
\end{equation}
Using these polyhedra, we can write down the mirror Calabi--Yau but as in
the $SO(7)$ case there is no choice of coordinates in which the fibration
over the lower sphere has a simple interpretation in terms of the number
of fundamentals and spinors. However, we can add the vector $v_9=(n-4,1,0,1)$
to the Newton polyhedron and use $x_8,x_9,x_3,x_5,x_6$ with $x_0=x_5x_6$
and $\zeta = (x_5/x_6)^{6+3n}$ to parametrize the Calabi--Yau
\begin{eqnarray}
W &=& \left( \zeta+a_0\frac{x_8^{4-n}x_9^{8-n}}{\zeta}\right) x_0^{6+3n}+
	2x_8 x_9^3x_0^{12}-x_8^2x_9+x_3^2+a_4 x_3x_8x_9x_0+a_5x_8^2x_9^4x_0^6\nl
	&&+a_6x_8^3x_9^5+a_7x_9^2x_0^{18}+a_8x_9x_0^{16}+a_9x_0^{14}+
	a_{10}x_3x_9x_0^9+a_{11}x_3x_0^7.
\end{eqnarray}
Near the singularity we can neglect the terms with coefficients
$a_{5,6}$. Substituting
\begin{eqnarray}
x_3 &=& x-\frac{1}{2}\left(a_{11}+ a_{10}x_9+a_4 x_8x_9\right) \nl
x_8 &=& y-\frac{1}{4}\left( a_{11}a_4+a_{10}a_{4} x_9 - 4 x_9^2\right)\\
x_9 &=& z\nonumber
\end{eqnarray}
into the defining equation of the Calabi--Yau gives
\begin{equation}
\label{so11curve}
W = \left( \zeta+a_0\frac{x_8^{4-n}z^{8-n}}{\zeta}\right)+
	x^2+z^5-zy^2+c_5z^4+c_4z^3+c_3z^2+c_2z+c_1+\cdots,
\end{equation}
where
\begin{equation}
x_8 = y-\frac{1}{8}\left( c_5^2-4c_4-4c_5z-8z^2\right).
\end{equation}
For $n=4$ we can integrate out $y$ trivially. Substituting $z\to z^2$, 
\eq{so11curve} reduces to the double cover version of the curve for $SO(11)$
with three fundamentals \cite{all,so2n1}. In the other two asymptotically
free cases $n=2,3$, we also obtain a curve but it is not hyperelliptic. 
Presumably we can obtain an exact solution for any number of massive
vectors and spinors by substituting
\begin{equation}
\left( \zeta+a_0\frac{x_8^{4-n}z^{8-n}}{\zeta}\right) \to
        \zeta+a_0\frac{1}{\zeta}z \prod_{i=1}^{N_s}(x_8-m^4_i)
                 \prod_{j=1}^{N_f}(z-m^2_j).
\end{equation}
Again, our solution passes the tests given in Sec.~\ref{iia}.

\section{Conclusions}
\label{iii}

We have obtained exact solutions to ${\cal N}=2$ supersymmetric $SO(N)$ gauge
theories for $N=10,12$ and $N=7,9,11$ with massless matter in the spinor and the
fundamental representation. We gave a description of the Coulomb branch of
these theories in terms of ALE spaces fibered over a sphere.

These solutions were obtained by compactifying type IIA string theory on
Calabi--Yau
threefolds with singular $K3$ fibers. The singularity type of the $K3$
determines the gauge group of the $d=4$ gauge theory and the duality to 
heterotic strings compactified on $K3\times T^2$ can be used to determine
the charged matter content of the theory. Mirror symmetry relates the
Calabi--Yau for type IIA compactification to a different Calabi--Yau that gives
rise to the same field theory when type IIB string theory is compactified on
it. The exact solutions can be extracted from this mirror Calabi--Yau.

This approach provides exact solutions for the gauge theories listed above
with specific matter contents. We proposed some generalizations of these
results to arbitrary numbers of massive spinors and vectors and verified
that our solutions are consistent under adjoint breaking and integrating out
matter fields. Unfortunately,
the list of asymptotically free $SO(N)$ theories with spinors is not exhausted
by the cases we have studied. For $SO(8)$ there is no toric description of
the corresponding type IIA and IIB Calabi--Yau manifolds and the higher rank
groups
$SO(N), N=13,14,15,16$ cannot be obtained from compactifying type IIA on
a Calabi--Yau threefold or conversely from breaking the adjoint of $E_8$
on the heterotic side. 

The results presented may ultimately provide some insights into how
to construct matter representations other than fundamentals and two index
tensors from branes. In principle it should be possible to find a brane
configuration corresponding to the theories we analyzed here by studying
an M-theory 5-brane wrapped on $R^4\times \Sigma$ where $\Sigma$ is the curve
encoding the gauge couplings on the Coulomb branch. 

Since our solutions agree with known field theory results, in the cases
where these are available, one can view the results of this paper as 
further confirmation of mirror symmetry and
the duality between type IIA and heterotic strings.

\acknowledgments
It is a pleasure to thank P.~Berglund, M.~Bershadsky, S.~Cherkis,
K.~Intriligator,
D.~Ramakrishnan, J.H.~Schwarz and especially A.~Van Proeyen for many helpful
conversations and E.~Gimon for amusing suggestions.
This work was supported in
part by the Department of Energy under Grant No.~DE-FG03-92-ER40701.

{\tighten

}

\end{document}